\magnification 1200
\centerline {\bf  Note on the Relativistic Thermodynamics of Moving Bodies}
\vskip 0.5cm
\centerline {by Geoffrey L. Sewell}
\vskip 0.3cm
\centerline {Department of Physics, Queen Mary University of London, Mile End Road, 
London E1 4NS, UK}
\vskip 0.2cm
\centerline {e-mail: g.l.sewell@qmul.ac.uk}
\vskip 1cm
\centerline {\bf Abstract}
\vskip 0.3cm
We employ a novel thermodynamical argument to show that, at the macroscopic level, 
there is no intrinsic law of temperature transformation under Lorentz boosts. This result 
extends the corresponding microstatistical one of earlier works to the purely macroscopic 
regime and signifies that the concept of temperature as an objective entity is restricted to 
the description of bodies in their rest frames. The argument on which this result is based 
is centred on the thermal transactions between a body that moves with uniform velocity 
relative to a certain inertial frame and a thermometer, designed to measure its 
temperature, that is held at rest in that frame. 
\vfill\eject
\centerline {\bf 1. Introduction and Discussion}
\vskip 0.3cm
Classical thermodynamics has been extended to the special relativistic regime in a 
number of different, logically consistent, ways, which have led to different formulae for 
the relationship between the temperature, $T_{0}$, of a body in a rest frame, $K_{0}$, 
and its temperature, $T$, in an inertial frame, $K$, that moves with velocity $v$ relative 
to $K_{0}$. To be specific, in the schemes of Planck [1] and Einstein [2],
$T=T_{0}(1-v^{2}/c^{2})^{1/2}$; whereas in those of Ott [3] and Kibble [4], 
$T=T_{0}(1-v^{2}/c^{2})^{-1/2}$; and in those of Landsberg [5], Van Kampen [6] and 
Callen and Horowitz [7], $T=T_{0}$, i.e. temperature is a scalar invariant. The 
relationships between the conventions and assumptions behind these different formulae 
has been lucidly exposited by Van Kampen  [6]. 
\vskip 0.2cm
In fact, all the above works were based exclusively on relativistic extensions of the first 
and second laws of classical thermodynamics. A different, quantum statistical, approach 
was introduced by Costa and Matsas [8] and by Landsberg and Matsas [9], who 
investigated the action of black body radiation on a monopole that moved with uniform 
velocity relative to the rest frame of the radiation and played the role of a thermometer or 
detector. The result they obtained was that the spectrum of the radiation, as registered by 
this detector, was non-Planckian, and therefore that it was only in a rest frame that the 
radiation had a well defined temperature.
\vskip 0.2cm
A much more general version of this result was obtained by the present author [10, 11], 
who showed that the coupling of a moving macroscopic quantum system, 
${\Sigma}_{0}$, to a fixed finite probe, ${\Sigma}$,  drives the latter to a terminal state 
that, generically, is non-thermal. This signifies that, at the microstatistical level, the 
concept of temperature, as measured by {\it any}, possibly microscopic, probe is 
restricted to systems in their rest frames. There remain, therefore, the open questions of 
whether the temperature of a moving body, as registered by macroscopic observables of a 
probe or thermometer, is well defined  and, if so, whether it transforms, under Lorentz 
boosts, according to some general law.
\vskip 0.2cm
These are the questions that we address in the present article by an argument based on the 
classical thermodynamics of  the composite, ${\Sigma}_{c}$, of two macroscopic bodies 
${\Sigma}$ and ${\Sigma}_{0}$, subject to the following conditions. ${\Sigma}_{0}$ 
is in thermal equilibrium at temperature $T_{0}$ in a rest frame $K_{0}$ and moves 
with uniform velocity $v$ relative to a frame $K$ in which ${\Sigma}$ is clamped at 
rest. Here again ${\Sigma}$ serves as a thermometer for ${\Sigma}_{0}$. We 
investigate whether the  coupling between ${\Sigma}$ and ${\Sigma}_{0}$ can drive 
${\Sigma}$ to equilibrium at a temperature $T$ that depends on $T_{0}$ and $v$ only: 
if so, $T$ would be interpreted as the temperature of the moving body ${\Sigma}_{0}$, 
relative to the frame $K$.  In fact, we show that there is no such model-independent 
temperature $T$. Hence, in the purely macroscopic picture, as well as in the quantum 
microstatistical one of Refs. [8]-[ 11], the concept of temperature as an objective entity is 
limited to bodies in their rest frames. 
\vskip 0.2cm
We formulate the thermodynamic description of 
${\Sigma}_{c}=({\Sigma}+{\Sigma}_{0})$ in Sec. 2, concluding that Section with the 
observation that its entropy can increase indefinitely and therefore that it cannot evolve 
into a true equilibrium state. This, however, does not preclude the possibility that 
${\Sigma}_{0}$ might drive ${\Sigma}$ into an equilibrium state, and in Sec. 3 we 
investigate this possibility for a specific tractable model in which ${\Sigma}$ and 
${\Sigma}_{0}$ interact via emission and absorption of radiation. This model is a variant 
of the one constructed by Van Kampen [6] for his treatment of heterotachic processes. 
We show that, for this model, ${\Sigma}$ is indeed driven into a thermal equilibrium 
state, but that the resultant temperature depends on variable parameters of this system.  
Accordingly we conclude in Sec. 4 that, since the temperature attained by ${\Sigma}$ is 
just that of the moving body ${\Sigma}_{0}$, as measured by a fixed thermometer, there 
is no intrinsic law of temperature transformations under Lorentz boosts. This result 
extands those of [8]-[11] from the microstatistical picture to the purely macroscopic one. 
\vskip 0.5cm
\centerline {\bf  2. The Thermodynamic Description}
\vskip 0.3cm 
Let ${\Sigma}_{0}$ be a macroscopic system that moves with velocity $v$ relative to an 
inertial frame $K$ and that is in equilibrium at temperature $T_{0}$ relative to a rest 
frame $K_{0}$. In order to formulate its thermodynamics relative to $K$, we consider 
the situation in which it is placed in diathermic interaction with a macroscopic probe, 
${\Sigma}$,  that is clamped at rest relative to $K$. We assume that the clamp is 
infinitely massive, and therefore immovable, and that its action on ${\Sigma}$ is 
adiabatic. Under these conditions, there is no thermal or mechanical exchange of energy, 
relative to $K$, between ${\Sigma}$ and the clamp. Further, we assume that the systems 
${\Sigma}$ and ${\Sigma}_{0}$ are spatially separated, so that they do not exchange 
energy by mechanical means.
\vskip 0.2cm
The transactions between ${\Sigma}_{0}$ and ${\Sigma}$ constitute a heterotachic 
process, as defined by Van Kampen [6], but with the crucial constraint that the 
momentum of ${\Sigma}$, relative to $K$, is held at the value zero. In this process, the 
energy relative to $K$ of the composite ${\Sigma}_{c}=({\Sigma}+{\Sigma}_{0})$ is 
conserved, but its momentum is not: any momentum received by ${\Sigma}$ is 
immediately discharged into the immovable clamp. 
\vskip 0.2cm
We assume that, although  both ${\Sigma}$ and ${\Sigma}_{0}$ are macroscopic, the 
former is of much smaller size than the latter in that, if  ${\Omega}$ and 
${\Omega}_{0}$ are dimensionless extensivity parameters (e.g. particles numbers) that 
provide measures of their respective sizes, then ${\Omega}_{0}>>{\Omega}>>1$. In 
order to sharpen our formulation, we take ${\Sigma}_{0}$ to be an infinite system, as in 
[10, 11], so  that ${\Omega}_{0}={\infty}$. Thus, ${\Sigma}_{0}$ serves as a thermal 
reservoir whose temperature and pressure remain constant during its transactions with 
${\Sigma}$. 
\vskip 0.2cm
We assume, for simplicity, that the energy $E$ and volume $V$ of ${\Sigma}$, relative 
to the rest frame $K$, constitute a complete set of its extensive thermodynamical 
variables\footnote*{A general quantum statistical characterisation of a complete set of 
extensive thermodynamical variables is provided in [12, Sec. 6.4].}. In fact, $V$ is 
merely constant during the transactions between this system and ${\Sigma}_{0}$ since, 
as stipulated above, no mechanical work is done on it relative to its rest frame. As for 
${\Sigma}_{0}$, we assume that its temperature $T_{0}$ and pressure ${\Pi}_{0}$, 
relative to $K_{0}$, together with its velocity $v$ relative to $K$, constitute a complete 
set of its intensive thermodynamic control variables. Finite changes from the equilibrium 
state of this system are given by increments $E_{0}$ and $P_{0}$ of its energy and 
momentum, respectively, relative to $K_{0}$. Hence, by Lorentz transformation, the 
increment in its energy relative to $K$ is 
$(1-v^{2}/c^{2})^{-1/2}(E_{0}+v.P_{0})$ and therefore the conservation of energy 
condition for ${\Sigma}_{c}$, relative to $K$, is .
$$E+{\gamma} (E_{0}+v.P_{0})=const.,\eqno(2.1)$$ 
where 
$${\gamma}=(1-v^{2}/c^{2})^{-1/2}.\eqno(2.2)$$
Note that it would be wrong to assume energy conservation relative to $K_{0}$, since 
energy in this frame is a linear combination of energy and momentum in $K$, and the 
clamping condition destroys the conservation of momentum of ${\Sigma}_{c}$ relative 
to the latter frame.
\vskip 0.2cm
The entropy of ${\Sigma}$ is a function $S$ of $E$ and $V$, which is jointly concave in 
its arguments [13, Sec. 1.10], and its value is Lorentz invariant [6; 14, Sec. 46]. The 
temperature $T$ of ${\Sigma}$ is related to $S$ by the standard formula
$$T^{-1}={{\partial}S(E,V)\over {\partial}E}.\eqno(2.3)$$ 
Since $K_{0}$ is a rest frame for ${\Sigma}_{0}$, the incremental entropy of this 
system, due to modification of its equilibrium state by changes $E_{0}$ and $P_{0}$ of 
its energy and momentum relative to this frame, is simply
$$S_{0}(E_{0})=T_{0}^{-1}E_{0}.\eqno(2.4)$$
The total entropy of the composite ${\Sigma}_{c}$, as measured relative to the specified 
equilibrium state of ${\Sigma}_{0}$, is just the sum of those of ${\Sigma}$ and 
${\Sigma}_{0}$, which, by Eqs. (2.1) and (2.4), is equal to 
$S(E,V)- T_{0}^{-1}({\gamma}^{-1}E+v.P_{0})$, plus a constant. Hence, defining
$${\tilde T}={\gamma}T_{0}\eqno(2.5)$$
and
$${\tilde S}(E,V)=S(E,V)-{\tilde T}^{-1}E,\eqno(2.6)$$
the entropy of ${\Sigma}_{c}$ is
$$S_{c}(E,V;P_{0})={\tilde S}(E,V)-T_{0}^{-1}v.P_{0}+const..\eqno(2.7)$$
\vskip 0.2cm 
We now note that it follows from Eq. (2.6) and the concavity of $S$ that ${\tilde S}$ is 
maximised at the value of $E$ for which ${\partial}S(E,V)/{\partial}E={\tilde T}^{-1}$
and that the resultant value of ${\tilde S}$ is the finite quantity given by  
$-{\tilde T}^{-1}$ times the Helmholtz free energy of ${\Sigma}$ at temperature 
${\tilde T}$ and volume $V$ [13, Sec.5.3]. On the other hand, the second term on the 
r.h.s. of Eq. (2.7) increases indefinitely with the modulus of $P_{0}$ when the direction 
of this excess momentum opposes that of $v$. Hence, $S_{c}$ has no finite upper bound 
and so we reach the following conclusion.
\vskip 0.3cm
(I) {\it Under the prescribed conditions, the composite system ${\Sigma}_{c}$ does not 
support any equilibrium state, as defined by the maximum entropy condition.}
\vskip 0.3cm
Of course this does not rule out the possibility that ${\Sigma}$ might be driven into a 
thermal state, with well defined temperature, as a result of its interaction with 
${\Sigma}_{0}$. In the following Section, we shall show that this possibility is realised 
by a tractable model, but that the resultant temperature varies with the parameters of the 
model. 
\vskip 0.5cm
\centerline {\bf 3. The Radiative Transfer Model}
\vskip 0.3cm
The model presented here is a variant of Van Kampen\rq s [6] system of two bodies that 
interact by radiation through a small hole in a metallic sheet placed between them. In the 
present context, these systems are the above described ones ${\Sigma}$ and 
${\Sigma}_{0}$. We assume that their respective boundaries facing the sheet are plane 
surfaces, $F$ and $F_{0}$, that are parallel both to it and to the velocity $v$. We assume 
that the sheet and the face $F_{0}$ are unbounded and that the sheet is at rest relative to 
$K$. Further, we assume that the hole is in the part of the sheet given by the orthogonal  
projection of $F$ onto it and that both the linear span of the hole and its distance from 
$F$ are negligibly small\footnote*{The distance of the hole from $F$ has to be so small 
in order to suppress end effects at the boundary of  that surface.} by comparison with its 
distance from the boundary of that face. 
\vskip 0.2cm
The modifications of Van Kampen\rq s model that we introduce here are the following.
\vskip 0.2cm\noindent
${\bullet}$  Only ${\Sigma}_{0}$, but not ${\Sigma}$, is a black body. We denote by 
$A({\omega})$ the absorption coefficient of ${\Sigma}$ for radiation of frequency 
${\omega}$. By Kirchoff\rq s law [15, Sec. 60], it is also the emission coefficient  of this 
system, and it necessarily lies in the interval [0,1]. 
\vskip 0.2cm\noindent
${\bullet}$ ${\Sigma}$ is clamped at rest in $K$.
\vskip 0.2cm\noindent
${\bullet}$ No radiation emanating from ${\Sigma}_{0}$ falls on the clamp: this can be 
achieved by placing ${\Sigma}$ between the hole and the clamp.
\vskip 0.3cm
{\bf 3.1. The Energy Exchanges.} Our treatment of the transactions between ${\Sigma}$ 
and ${\Sigma}_{0}$ will be basd on a calculation of the increment in the energy, 
${\Delta}E$, of ${\Sigma}$ relative to $K$  in time ${\Delta}t$. Evidently this may be 
expressed in the form
$${\Delta}E={\Delta}E_{2}-{\Delta}E_{1},\eqno(3.1)$$
where ${\Delta}E_{1}$ (resp. ${\Delta}E_{2}$) is the energy transferred from 
${\Sigma}$ to ${\Sigma}_{0}$ (resp. ${\Sigma}_{0}$ to ${\Sigma}$) in that time. 
These energy transfers are achieved by leaks of the radiations emanating from 
${\Sigma}$ and ${\Sigma}_{0}$ through the hole in the metallic sheet. Since both the 
linear span of the hole and its distance from $F$ are negligible by comparison with its 
distance from the boundary of $F$, we may assume, for the purpose of calculating 
${\Delta}E$, that the face $F$, as well as $F_{0}$, is infinitely extended. We denote by 
${\Gamma}$ (resp. ${\Gamma}_{0}$) the region bounded by $F$ (resp. $F_{0}$) and 
the sheet. Thus ${\Gamma}$ and ${\Gamma}_{0}$ are are filled with the thermal 
radiation emanating from ${\Sigma}$ and ${\Sigma}_{0}$, respectively, as modified by 
the leakages through the hole.
\vskip 0.2cm 
In order to calculate ${\Delta}E_{1}$, we first note that the energy density of the pencil 
of radiation in ${\Gamma}$ that lies in the infinitesimal frequency range 
$[{\omega},{\omega}+d{\omega}]$ and whose direction lies in a solid angle 
$d{\Omega}$ is 
$A({\omega}){\omega}^{3}[{\rm exp}({\hbar}w/kT)-1]^{-1}d{\omega}d{\Omega}$, 
times a universal constant. Hence, denoting the area of the hole by ${\Delta}a$,  the 
energy transferred by this pencil of radiation from ${\Gamma}$ to ${\Gamma}_{0}$ in 
time ${\Delta}t$ is 
$$C{\Delta}a{\Delta}tA({\omega}){\omega}^{3}
[{\rm exp}({\hbar}{\omega}/kT)-1]^{-1}{\rm cos}({\psi})d{\omega}d{\Omega},$$
where $C$ is a universal constant and ${\psi}$ is the angle between the pencil and the 
outward drawn normal to the sheet. It is convenient to express $d{\Omega}$ and 
${\rm cos}({\psi})$ in terms of  spherical polar coordinates ${\theta} \ ({\in}[0,{\pi}])$ 
and  ${\phi} \ ({\in}[-{\pi}/2,{\pi}/2])$, where the former is the angle between the pencil 
and the direction of $v$ and the latter is the azimuthal angle  of rotation of the pencil 
about the line of $v$. Specifically,
$$d{\Omega}={\rm sin}({\theta})d{\theta}d{\phi} \ {\rm and} \ {\rm cos}({\psi})=
{\rm sin}({\theta}){\rm cos}({\phi})$$ 
and therefore the above expression for the energy transferred across the hole from 
${\Gamma}$ may be re-expressed as  
$$C{\Delta}a{\Delta}tA({\omega}){\omega}^{3}
[{\rm exp}({\hbar}{\omega}/kT)-1]^{-1}{\rm sin}^{2}({\theta}){\rm cos}({\phi})
d{\omega}d{\theta}d{\phi},\eqno(3.2)$$
Since ${\Sigma}_{0}$ is a black body, the total energy ${\Delta}E_{1}$, relative to 
$K$, that is transferred from ${\Sigma}$ to ${\Sigma}_{0}$ in time ${\Delta}t$ is 
obtained by integration of this quantity over the ranges $[0,{\infty}]$ for ${\omega}, \ 
[0,{\pi}]$ for ${\theta}$ and $[-{\pi}/2,{\pi}/2]$ for ${\phi}$. Thus 
$${\Delta}E_{1}=C{\Phi}(T){\Delta}a{\Delta}t,\eqno(3.3)$$
where
$${\Phi}(T)={\pi}\int_{0}^{\infty}d{\omega} A({\omega}){\omega}^{3}
[{\rm exp}({\hbar}{\omega}/kT)-1]^{-1}.\eqno(3.4)$$
Here there is the tacit mathematical assumption that the function $A$ is measurable: 
otherwise the integral in Eq. (3.4) would not be well defined. However, from the physical 
standpoint, this assumption is very mild, as it is satisfied if the function $A$ is piecewise 
continuous. It follows from Eq. (3.4) that ${\Phi}(T)$  is a continuous and monotonically 
increasing function of $T$ whose range is $[0,{\infty}]$. 
\vskip 0.2cm
The calculation of ${\Delta}E_{2}$ proceeds along similar lines, with modifications due 
to the motion of ${\Sigma}_{0}$ relative to $K$. To effect this calculation we first note 
that the radiation emanating from the black body ${\Sigma}_{0}$ is Planckian, and 
therefore isotropic, relative to $K_{0}$.We then define ${\omega}_{0}, \ {\theta}_{0}$ 
and ${\phi}_{0}$ to be the natural counterparts of ${\omega}, \ {\theta}$ and ${\phi}$, 
respectively, for the description of ${\Sigma}_{0}$ relative to $K_{0}$, and we denote 
by ${\cal P}_{0}$ the pencil of radiation emanating from ${\Sigma}_{0}$ for which 
these variables lie in the infinitesimal ranges 
$[{\omega}_{0},{\omega}_{0}+d{\omega}_{0}], \ 
[{\theta}_{0},{\theta}_{0}+d{\theta}_{0}]$ and 
$[{\phi}_{0},{\phi}_{0}+d{\phi}_{0}]$. We then note that  ${\Delta}a{\Delta}t$ is 
Lorentz invariant, i.e. it is equal to the product of the counterparts ${\Delta}a_{0}$ and 
${\Delta}t_{0}$ of ${\Delta}a$ and ${\Delta}t$ relative to the frame $K_{0}$. It now 
follows by simple analogy with the derivation of  (3.2) that the energy, relative to 
$K_{0}$, that is transferred by this pencil through the hole from ${\Gamma}_{0}$ to 
${\Gamma}$ in time ${\Delta}t_{0}$ is given by the canonical analogue of the 
expression (3.2), but with the term $A({\omega})$ omitted, since ${\Sigma}_{0}$ is a 
black body. Hence, in view of the Lorentz invariance of ${\Delta}a{\Delta}t$, the energy 
relative to $K_{0}$ transmitted by the pencil ${\cal P}_{0}$ through the hole in time 
${\Delta}t_{0}$ is
$$C{\Delta}a{\Delta}t{\omega}_{0}^{3}
[{\rm exp}({\hbar}{\omega}_{0}/kT_{0})-1]^{-1}{\rm sin}^{2}({\theta}_{0})
{\rm cos}({\phi}_{0})d{\omega}_{0}d{\theta}_{0}d{\phi}_{0}.$$
Correspondingly, the component parallel to $v$ of the momentum of ${\cal P}_{0}$, 
relative to $K_{0}$, that is transferred from ${\Gamma}_{0}$ to ${\Gamma}$ in time 
${\Delta}t_{0}$ is just $c^{-1}{\rm cos}({\theta}_{0})$ times this quantity. Hence, by 
Lorentz transformation, the energy of this pencil, relative to $K$, that is transferred to 
${\Sigma}$ in time ${\Delta}t$ is     
$$C{\gamma}{\Delta}a{\Delta}t{\omega}_{0}^{3}
[{\rm exp}({\hbar}{\omega}_{0}/kT_{0})-1]^{-1}\bigl(1+({\vert}v{\vert}/c)
{\rm cos}({\theta}_{0})\bigr){\rm sin}^{2}({\theta}_{0})
{\rm cos}({\phi}_{0})d{\omega}_{0}d{\theta}_{0}d{\phi}_{0}.\eqno(3.5)$$
Moreover, in view of the relativistic Doppler effect [14, Sec. 6], the frequency of this 
radiative pencil, relative to $K$, is 
$${\omega}={\gamma}\bigl(1+({\vert}v{\vert}/c)
{\rm cos}({\theta}_{0})\bigr){\omega}_{0}.\eqno(3.6)$$
Therefore, as viewed in $K$, the energy transferred by the pencil ${\cal P}_{0}$  from 
${\Sigma}_{0}$ to ${\Sigma}$  in time ${\Delta}t$ is just ${\gamma}$ times the 
expression (3.5), but with ${\omega}_{0}$ replaced by 
${\gamma}^{-1}\bigl(1+({\vert}v{\vert}/c){\rm cos}
({\theta}_{0})\bigr)^{-1}{\omega}$. Moreover, the resultant energy absorbed by 
${\Sigma}$ from the pencil is just the absorption coefficient $A({\omega})$ times this 
quantity. The total energy ${\Delta}E_{2}$ absorbed by ${\Sigma}$ in time ${\Delta}t$ 
is then obtained by integration and takes the form
$${\Delta}E_{2}=C{\Phi}_{0}(T_{0}){\Delta}a{\Delta}t,\eqno(3.7)$$
where
$${\Phi}_{0}(T_{0})=2{\gamma}^{-3}
\int_{0}^{\infty}d{\omega}\int_{0}^{\pi}d{\theta}_{0}
A({\omega}){\omega}^{3}{\rm sin}^{2}({\theta}_{0}){\times}$$
$$\bigl(1+({\vert}v{\vert}/c){\rm cos}({\theta}_{0})\bigr)^{-3}
\bigl[{\rm exp}\bigl(({\hbar}{\omega}/{\gamma}kT)\bigl(1+({\vert}v{\vert}/c)
{\rm cos}({\theta}_{0})\bigr)^{-1}\bigr)-1\bigr]^{-1}.\eqno(3.8)$$
It follows immediately from this formula that ${\Phi}_{0}$ is a continuous, 
monotonically increasing function of $T_{0}$ whose range is $[0,{\infty})$.
\vskip 0.2cm
We now infer from Eqs. (3.1), (3.3) and (3.7) that the net energy increment in the energy 
of ${\Sigma}$, relative to $K$, in time ${\Delta}t$ is 
$${\Delta}E=C [{\Phi}_{0}(T_{0})-{\Phi}(T)] {\Delta}a{\Delta}t .\eqno(3.9)$$
Hence, passing to the limit ${\Delta}t{\rightarrow}0$, the rate of change of the energy 
$E$ of ${\Sigma}$ is
$${dE\over dt}= C [{\Phi}_{0}(T_{0})-{\Phi}(T)] {\Delta}a.\eqno(3.10)$$
\vskip 0.3cm
{\bf 3.2. Evolution to the Equilibrium Temperature of ${\Sigma}$.} Since the functions 
${\Phi}$ and ${\Phi}_{0}$ are continuous and monotonically increasing, with range 
$[0, \ {\infty})$, it follows from Eq. (3.10) that there is precisely one value, ${\overline 
T}$, of $T$ for which $E$ is stationary. Thus ${\overline T}$ is determined by the 
equation
$${\Phi}({\overline T})={\Phi}_{0}(T_{0}).\eqno(3.11)$$
Moreover, since ${\Phi}_{0}$, as well as ${\Phi}$, increases monotonically and 
continuously with its argument, this formula implies that ${\overline T}$ is an increasing 
function of $T_{0}$.
\vskip 0.2cm
In order to show that the temperature of ${\Sigma}$ evolves irreversibly to the value 
${\overline T}$, we introduce the free energy function
$$F(E,V)=E-{\overline T}S(E,V)\eqno(3.12)$$
and infer from Eqs. (2.3) and (3.10) that
$${d\over dt}F(E,V)=C[1-{\overline T}/T]
[{\Phi}_{0}(T_{0})-{\Phi}(T)] {\Delta}a$$
and consequently, by Eq. (3.11), that
$${d\over dt}F(E,V)=C [1-{\overline T}/T]
[{\Phi}({\overline T})-{\Phi}(T)] {\Delta}a.\eqno(3.13)$$
Since ${\Phi}$ is a continuous monotonically increasing function of temperature it 
follows immediately from this equation that $dF/dt$ is negative except at 
$T={\overline T}$, where it is zero. This leads us to the following result.
\vskip 0.3cm
(II) {\it  $F$ serves as a Lyapounov function whose monotonic decrease with time 
ensures that the temperature of ${\Sigma}$ evolves irreversibly to a stable terminal value 
${\overline T}$, which is the temperature of  the moving system ${\Sigma}_{0}$, as 
registered by the thermometer fixed in $K$. Moreover, as noted following Eq. (3.9), this 
temperature is an increasing function of $T_{0}$. }
\vskip 0.3cm
{\bf 3.3. Dependence of ${\overline T}$ on the Parameters of the Model.}We now 
remark that, by Eqs. (3.4) and (3.8), the functions ${\Phi}$ and ${\Phi}_{0}$ depend on 
the form of the absorption coefficient $A({\omega})$, and therefore, by Eq. (3.11), so 
too does the temperature ${\overline T}$. In order to establish that this dependence is 
non-trivial, we consider the case where $A({\omega})$ is unity when ${\omega}$ lies in 
a narrow interval $[f,f+{\Delta}f]$ and is otherwise zero. In this case, it follows from 
Eqs. (3.4), (3.8) and (3.11) that
$$[{\rm exp}({\hbar}f/k{\overline T})-1]^{-1}=2{\pi}^{-1}{\gamma}^{-3}
\int_{0}^{\pi}d{\theta}_{0}{\rm sin}^{2}({\theta}_{0}){\times}$$
$$\bigl(1+({\vert}v{\vert}/c){\rm cos}({\theta}_{0})\bigr)^{-3}
\bigl[{\rm exp}\bigl(({\hbar}f/{\gamma}kT_{0})
\bigl(1+({\vert}v{\vert}/c){\rm cos}({\theta}_{0})\bigr)^{-1}\bigr)-1\bigr]^{-1}.
\eqno(3.14)$$
In order to establish that ${\overline T}$ depends non-trivially on the frequency $f$, i.e. 
that it is not just a constant, we show ${\overline T}$ tends to different limits as $f$  
tends to zero and infinity. Thus, in the case of small $f$, we may approximate the 
quantities in the square brackets on the left and right hand sides of Eq. (3.14) by the 
exponents occurring there. Thus we find that
$${\overline T}{\rightarrow}2{\pi}^{-1}{\gamma}^{-2}T_{0}\int_{0}^{\pi}
d{\theta}_{0}{\rm sin}^{2}({\theta}_{0})
\bigl(1+({\vert}v{\vert}/c){\rm cos}({\theta}_{0})\bigr)^{-2} \ {\rm as} \ 
f{\rightarrow}0.\eqno(3.15)$$
On the other hand, for large $f$, we may discount the terms $-1$ in the square brackets 
on both sides of Eq. (3.14), thereby obtaining the formula  
$${\rm exp}(-{\hbar}f/k{\overline T})=2{\pi}^{-1}{\gamma}^{-3}{\times}$$
$$\int_{0}^{\pi}d{\theta}_{0}{\rm sin}^{2}({\theta}_{0})
\bigl(1+({\vert}v{\vert}/c){\rm cos}({\theta}_{0})\bigr)^{-3}
{\rm exp}\bigl(-({\hbar}f/{\gamma}kT_{0})
\bigl(1+({\vert}v{\vert}/c){\rm cos}({\theta}_{0})\bigr)^{-1}\bigr).$$
For large $f$, the r.h.s. of this equation is dominated by the exponential term occurring 
therein and its logarithm reduces to the maximum value of the exponent for 
${\theta}_{0}{\in}[0,{\pi}]$. Hence, using Eq. (2.2), we find that
$${\overline T}{\rightarrow}T_{0}
\Bigl({{1+{\vert}v{\vert}/c}\over {1-{\vert}v{\vert}/c}}\Bigr)^{1/2} \  {\rm as} \ 
f{\rightarrow}{\infty}.\eqno(3.16)$$
This limit is evidently different from that of Eq. (3.15), since it follows easily from Eq. 
(2.2) that the latter limit is equal to $T_{0}\bigl(1+O(v^{2}/c^{2})\bigr)$. Hence the 
temperature ${\overline T}$ must be a nontrivial, i.e. non-constant, function of $f$. It 
follows that this temperature depends on the parameter of the model and therefore we 
arrive at the following general conclusion.
\vskip 0.3cm
(III) {\it According to the purely macroscopic picture, there is no intrinsic law of 
temperature transformations under Lorentz boosts. }
\vskip 0.5cm
\centerline {\bf 4. Conclusion}
\vskip 0.3cm
Our essential results are encapsulated by the assertions (I) of Sec. 2 and (II) and (III) of 
Sec. 3. The first of these is that, under the prescribed conditions, the composite of 
${\Sigma}$ and ${\Sigma}_{0}$ cannot evolve to an equilibrium state, as given by the 
maximum entropy condition. However, as in the case of Sec. 3, where these systems 
interact via radiative transfer, their coupling can drive ${\Sigma}$ into an equilibrium 
state whose temperature ${\overline T}$ varies not only with $T_{0}$ but also with the 
parameters of the  thermometer ${\Sigma}$. From this we conclude that in the purely 
macroscopic picture, as in the microstatistical one of [8]-[11],  there is no intrinsic law of 
temperature transformation under Lorentz boosts.
\vskip 0.5cm
{\bf  Acknowledgment.} The author wishes to thank a referee for correcting some 
mistakes in an earlier draft of this article.
\vskip 0.5cm 
\centerline {\bf References}
\vskip 0.3cm\noindent 
[1] Planck M.1907 {\it Sitzber. K1. Preuss. Akad. Wiss.}, 542
\vskip 0.2cm\noindent
[2] Einstein A. 1907 {\it Jahrb. Radioaktivitaet Elektronik} {\bf 4}, 411
\vskip 0.2cm\noindent
[3] Ott H. 1963 {\it Z. Phys.} {\bf 175}, 70
\vskip 0.2cm\noindent
[6] Kibble T. W. B. 1966 {\it Nuov. Cim.} {\bf 41 B}, 72
\vskip 0.2cm\noindent
[5] Landsberg P. T. 1966 {\it Nature} {\bf 212}, 571 ;
 \ \ and 1967 {\it Nature} {\bf 214}, 903
\vskip 0.2cm\noindent
[6] Van Kampen N. G. 1968 {\it Phys. Rev.} {\bf 173}, 295
\vskip 0.2cm\noindent
[7] Callen H. and Horowitz G. 1971 {\it Am. J. Phys.} {\bf 39}, 938
\vskip 0.2cm\noindent
[8] Costa S. S. and Matsas G. E. A. 1995 {\it Phys. Lett. A} {\bf 209}, 155.
\vskip 0.2cm\noindent
[9] Landsberg P. T. and Matsas G. E. A. 1996 {\it Phys. Lett. A} {\bf 223}, 401
\vskip 0.2cm\noindent
[10]  Sewell G. L. 2008 {\it J. Phys. A: Math. Theor.} {\bf 41}, 382003
\vskip 0.2cm\noindent
[11] Sewell G. L. 2009 {\it Rep. Math. Phys.} {\bf 64}, 285
\vskip 0.2cm\noindent
[12] Sewell G. L. 2002 {\it Quantum Mechanics and its Emergent Macrophysics} 
(Princeton NJ: Princeton University Press)
\vskip 0.2cm\noindent
[13]  Callen H. B. 1987 {\it Thermodynamics and an Introduction to Thermostatistics} 
(John Wiley; New York)
\vskip 0.2cm\noindent
[14] Pauli W. 1958 {\it Theory of Relativity}, (London: Pergamon)
\vskip 0.2cm\noindent
[15] Landau L. D. and Lifshitz E. M. 1959 {\it Statistical Physics} (Pergamon; London, 
Paris)
\end